\documentclass[%
 reprint,
 amsmath,amssymb,
 aps,
]{revtex4-2}

\usepackage{graphicx}
\usepackage{dcolumn}
\usepackage{bm}
\usepackage[colorlinks=true,bookmarks=false,citecolor=blue,urlcolor=blue]{hyperref}

\begin{document}

\preprint{APS/123-QED}

\title{High-efficiency high-NA metalens designed by maximizing the efficiency limit}

\author{Shiyu Li}
\author{Ho-Chun Lin}
\author{Chia Wei Hsu}%
\email{cwhsu@usc.edu}
\affiliation{%
Ming Hsieh Department of Electrical and Computer Engineering, University of Southern California, Los Angeles, CA 90089, USA
}%

\begin{abstract}
Theoretical bounds are commonly
used to assess the limitations of photonic design.
Here we introduce a more active way to use theoretical bounds, integrating them into part of the design process and identifying optimal system parameters that maximize the efficiency limit itself.
As an example, we consider wide-field-of-view high-numerical-aperture metalenses, which can be used for high-resolution imaging in microscopy and endoscopy, but no existing design has achieved a high efficiency.
By choosing aperture sizes to maximize an efficiency bound, setting the thickness according to a thickness bound, and then performing inverse design, we 
come up with
high-numerical-aperture (NA = 0.9) metalens designs with record-high 98\% transmission efficiency and 92\% Strehl ratio across all incident angles within a 60$^{\circ}$ field of view, reaching the maximized bound.
This maximizing-efficiency-limit approach applies to any multi-channel system and can help a wide range of optical devices reach their
highest possible performance.
\end{abstract}

\maketitle

\section{Introduction}\label{sec:intro}

With the rapid advance of nanophotonic design and fabrication, there has been increasing interests in exploring the ultimate limit of device performance.
In recent years, theoretical bounds have been proposed for the
operational bandwidth~\cite{shrestha2018broadband,presutti2020focusing,engelberg2021achromatic,shastri2022bandwidth,shastri2022extent}, efficiency ({\it e.g.}, reflection, transmission, absorption)~\cite{estakhri2016wave,asadchy2016perfect,arbabi2017fundamental,miller2022towards,li2023transmission}, coupling strengths~\cite{miller2019waves,kuang2022bounds}, energy concentration~\cite{zhang2019scattering,shim2020maximal}, and the device thickness~\cite{li2022thickness,2023_Miller_Science}.
Conventionally, these bounds are used to assess how close a design approaches fundamental limits~\cite{chao2022physical}. 
One can then set realistic targets and avoid futile attempts on trying to design things that are theoretically impossible.
While understanding the limit is important, it would be desirable if theoretical bounds can also play a more active role in improving the design and in pushing the limit itself.

\begin{figure}[t]
\centering
\includegraphics[width=0.33\textwidth]{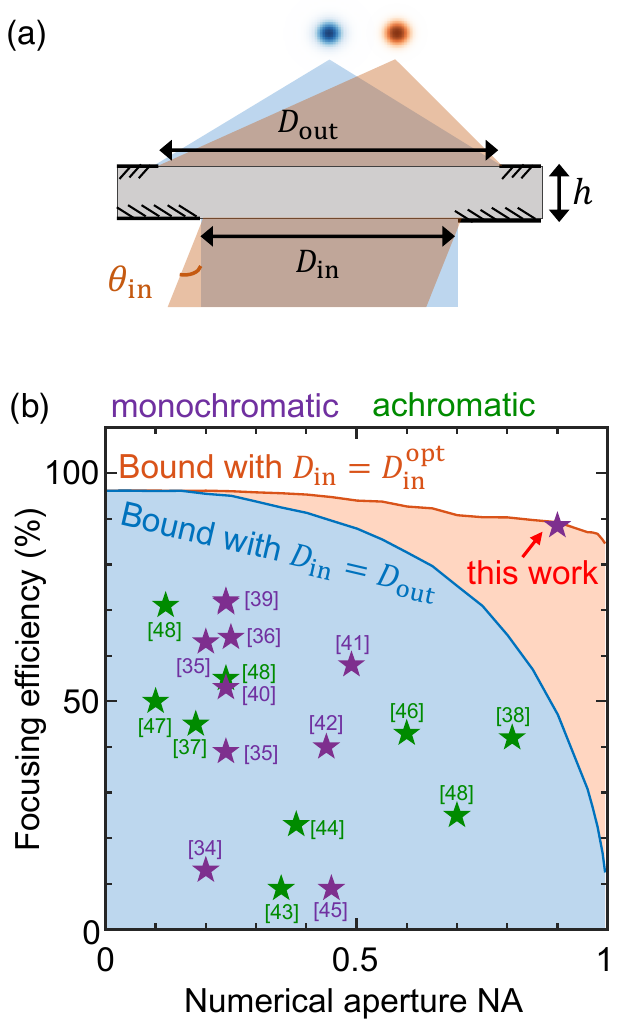}
\caption{
(a) Schematic of a high-NA wide-field-of-view metalens, with $D_{\rm in}$ ($D_{\rm out}$), $\theta_{\rm in}$, and $h$ denoting the input (output) aperture diameter, incident angle, and lens thickness.
(b) Comparison of the focusing efficiency in existing wide-FOV metalens designs and in this work
(purple for monochromatic, green for achromatic).
Details are listed in Supplementary Tab.~S1.
Solid lines show the theoretical limit~\cite{li2023transmission} with $D_{\rm in}=D_{\rm out}$ and with the optimal $D_{\rm in} = D_{\rm in}^{\rm opt}$ used here.
}
\label{fig:Fig1}
\end{figure}

An important application that calls for further progress is designing metalenses with a high numerical aperture (NA) over a wide field of view (FOV) [Fig.~\ref{fig:Fig1}(a)],
which can be used for high-resolution imaging in microscopy, endoscopy, and lithography.
Metalenses that offer diffraction-limited resolution at NA $\ge$ 0.9 have been proposed~\cite{arbabi2015subwavelength,chen2017immersion,liang2018ultrahigh,paniagua2018metalens,mohammad2018broadband,chung2020high,liu2020high,sang2022toward,wang2022manipulation,yang2023high,xue2023fullwave}, with focusing efficiencies exceeding 70\% reached through inverse design~\cite{byrnes2016designing,phan2019high,banerji2019imaging}.
However, these metalenses are only designed to operate at the normal incidence, 
and their focusing quality deteriorates sharply at oblique incidence~\cite{kalvach2016aberration,lin2022fast}, severely limiting the angular FOV.
There are multiple strategies to broaden the FOV while preserving diffraction-limited focusing, including using an aperture to separate different incident angles onto different parts of the metasurface~\cite{engelberg2020near,shalaginov2020single,fan2020ultrawide,yang2021design,qi2022broadband,yang2023understanding,lin2023wide}, using a doublet~\cite{arbabi2016miniature,groever2017meta,tang2020achromatic,kim2020doublet,li2021super,huang2021achromatic,jang2023dispersion}, or stacking multiple structural layers~\cite{lin2021computational}.
However, there is ample room for improvement in the efficiency, as summarized in Fig.~\ref{fig:Fig1}(b) and Supplementary Tab.~S1.
For NA $\ge 0.5$, no existing design achieved a focusing efficiency above 50\% over a wide FOV.

To identify and to reach the highest possible performance, here we propose to integrate theoretical bounds into part of the design process.
In the first step, we consider a recently introduced bound on the channel-averaged transmission efficiency~\cite{li2023transmission}, which depends on the sizes of the input and output apertures [$D_{\rm in}$ and $D_{\rm out}$ in Fig.~\ref{fig:Fig1}(a)]. 
Given any output aperture $D_{\rm out}$ of interest, we identify the optimal input aperture size $D_{\rm in}^{\rm opt}$ that maximizes the efficiency bound.
In the second step, we set the thickness $h$ based on a recently introduced bound on the minimal device thickness~\cite{li2022thickness}. 
In the third step, we perform inverse design using the optimal $D_{\rm in}$ and the minimal $h$.
Applying this approach to
a metalens with NA = 0.9 with a 60$^{\circ}$ FOV,
we achieve a transmission efficiency of 98\% and Strehl ratio of 92\% across the full FOV, reaching the maximized efficiency limit. 
Previous inverse designs attained significantly lower efficiencies because without the guidance from theoretical bounds, they invariably adopted suboptimal geometry parameters.
Both bounds we employ~\cite{li2023transmission,li2022thickness} follow from the target functionality of any linear multi-channel optical system, so the proposed strategy here is general, not limited to metalens designs.

\section{Numerical optimization method}\label{sec:method}
We first describe our inverse design methodology. We employ free-form topology optimization~\cite{bendsoe2003topology,2012_Miller_thesis,molesky2018inverse,fan2020freeform}, 
allowing non-intuitive permittivity profiles. 
We have made our code open-source, available at Ref.~\cite{inverse_design_code}.
We consider the transverse magnetic waves $(E_x, H_y, H_z)(y,z)$ of a system invariant in the $x$ dimension. 
Given the $a$-th incident angle $\theta_{\rm in}^a$ within the FOV, the desired focal spot position is ${\bf r}_{\rm f}^a = (y=f\tan{\theta_{\rm in}^a},z=f)$ where $f$ is the focal length.
We quantify the focusing quality by the ratio between the intensity of the actual design at ${\bf r}_{\rm f}^a$ and the focal spot intensity for an ideal lens (\textit{i.e.}, perfectly transmitting and free of aberrations), as
\begin{equation}
    I_a = \frac{|E_x(y=f\tan{\theta_{\rm in}^a},z=f;\theta_{\rm in}^a)|^2}{{\rm max}_y|E_{x,{\rm ideal}}(y,z=f;\theta_{\rm in}^a)|^2}={\rm SR}_a \cdot T_a,
    \label{eq:FoM}
\end{equation}
which is the product of the Strehl ratio (SR) and the transmission efficiency $T$ at incident angle $\theta_{\rm in}^a$.
The SR~\cite{2012_lens_design,lin2021computational} is defined as the actual intensity at the focal spot divided by the peak intensity of an ideal diffraction-limited focus normalized by the total transmitted power of the metalens.
The transmission efficiency $T$ is the total transmitted power divided by the incident power.

We map out the focal intensity as a function of the incident angle through the generalized transmission matrix $\mathbf{t}=\mathbf{CA}^{-1}\mathbf{B}$ of the metalens~\cite{lin2022fast}.
Here, matrix $\mathbf{A}=-\nabla^2-(2\pi/\lambda)^2\varepsilon_{\rm r}(\mathbf{r})$ is the discretized Maxwell differential operator on $E_x(y,z)$ at wavelength $\lambda$ for the metalens structure defined by its relative permittivity profile $\varepsilon_{\rm r}(\mathbf{r})$.
The $a$-th column of the $M$-column input matrix $\mathbf{B}=[\mathbf{B}_1,\cdots,\mathbf{B}_M]$ contains the source profile that generates an incident plane wave at the $a$-th incident angle $\theta_{\rm in}^a$ within the input aperture.
The $b$-th row of the $M$-row output projection matrix $\mathbf{C}=[\mathbf{C}_1; \cdots;\mathbf{C}_M]$ performs angular spectrum propagation~\cite{goodman2005introduction} that propagates $E_x$ on the output surface of the metalens to position ${\bf r}_{\rm f}^b$ on the focal plane.
Then, the diagonal elements $t_{aa}$ are the field amplitudes $E_x({\bf r}_{\rm f}^a;\theta_{\rm in}^a)$ at the focal spots in Eq.~(\ref{eq:FoM}), for all incident angles $\{\theta_{\rm in}^a\}$ within the FOV of interest.
More details are given in Supplementary Secs.~1--2.

To reach optimal $I_a ={\rm SR}_a \cdot T_a$ for all incident angles over the FOV, we maximize the worst-case ${\rm min}_a\ I_a$ within the FOV.
To make the objective function differentiable, we cast the problem into an equivalent epigraph form~\cite{boyd2004convex} using a dummy variable $g$,
\begin{equation}
\begin{split}
    & {\rm max}_{g,\varepsilon_{\rm r}}\ \ \ g, \\
    & \text{subject to }g \leq I_a(\varepsilon_{\rm r}).
\end{split}
\label{eq:opt_problem}
\end{equation}
We perform the optimization using the gradient-based method of moving asymptotes~\cite{svanberg2002class} implemented in the open-source package NLopt~\cite{NLopt}.
Under finite-difference discretization with grid size $\Delta x$,
the gradient of $I_a$ with respect to the permittivity profile $\varepsilon_{\rm r}$ of the metalens is calculated via the adjoint method (see Supplementary Sec.~3) as
\begin{equation}
    \frac{dI_a}{d\varepsilon_{\rm r}({\bf r})} = \frac{2k_0^2\Delta x^2 {\rm Re}\left[ (\mathbf{C}_a\mathbf{A}^{-1}\mathbf{B}_a)^{*} (\mathbf{A}^{-1}\mathbf{C}_a^{\rm T} \circ \mathbf{A}^{-1}\mathbf{B}_a) \right]({\bf r})}{{\rm max}_y|E_{x,{\rm ideal}}(y,z=f;\theta_{\rm in}^a)|^2},
    \label{eq:grad}
\end{equation}
where $k_0=2\pi/\lambda$ is the free-space wave vector; operators $\circ$, $^*$, and $^{\rm T}$ stand for the element-wise Hadamard product, complex conjugation, and vector transpose respectively.
We perform the computation with the open-source multi-channel Maxwell solver MESTI~\cite{MESTI} using single-precision arithmetic.

\begin{figure}[t]
\centering
\includegraphics[width=0.4\textwidth]{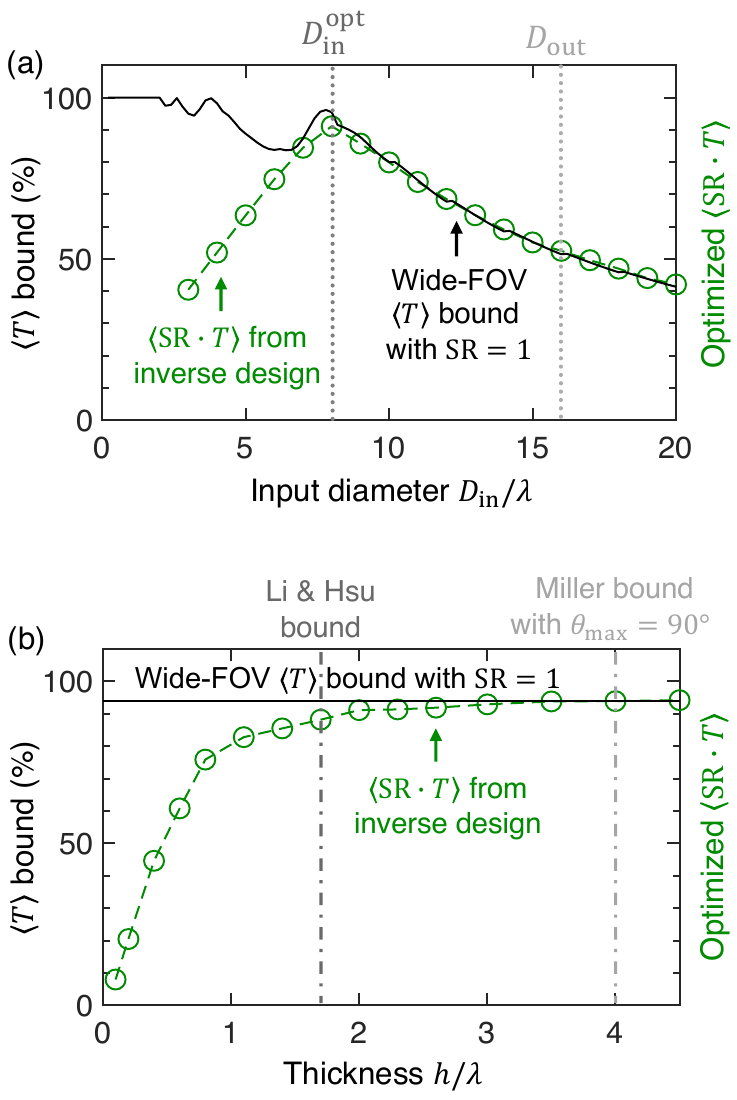}
\caption{Validation of theoretical bounds on the channel-averaged total transmission efficiency $T$ and on the device thickness $h$ for a high-NA wide-field-of-view metalens.
Black solid lines are bounds on the angle-averaged transmission efficiency $\langle T \rangle$ from Ref.~\cite{li2023transmission}, assuming a perfect Strehl ratio ${\rm SR}=1$ across all incident angles within the FOV.
Vertical dot-dashed lines in (b) indicate the thickness bounds from Li and Hsu~\cite{li2022thickness} and from Miller~\cite{2023_Miller_Science}.
Green circles are the angle-averaged $\langle {\rm SR} \cdot T \rangle$ from gray-scale topology optimizations.
Lens parameters: numerical aperture NA = 0.9, output aperture diameter $D_{\rm out}=16\lambda$, field of view FOV = 60$^{\circ}$, refractive index range = $[1,2]$.
Thickness $h = 2\lambda$ in (a), and input diameter $D_{\rm in} = D_{\rm in}^{\rm opt} = 8\lambda$ in (b).}
\label{fig:Fig2}
\end{figure}

\begin{figure*}[tb]
\centering
\includegraphics[width=0.8\textwidth]{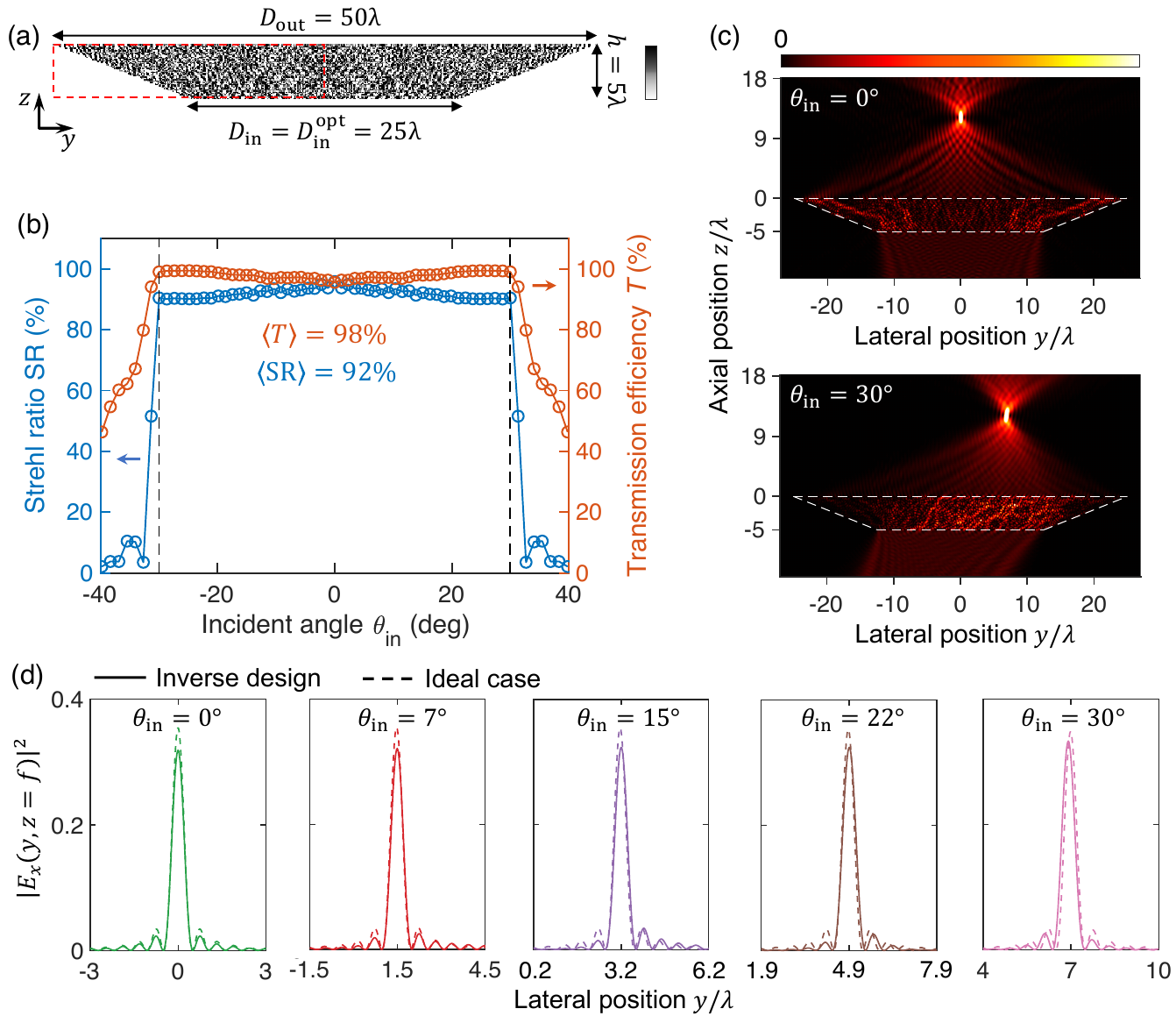}
\caption{An inverse-designed high-NA wide-FOV metalens guided by theoretical bounds.
(a) Optimized metalens structure, with color indicating the refractive index (between 1.0 and 2.0).
(b) The Strehl ratio SR and transmission efficiency $T$ of the optimized metalens as a function of the incident angle $\theta_{\rm in}$. Vertical dashed lines mark the FOV. 
(c) Intensity profiles $|E_x(y,z)|^2$ for incident angles $\theta_{\rm in}=0^{\circ}$ and 30$^{\circ}$. White dashed lines mark the boundary of the metalens.
(d) Intensity profiles at the focal plane $z=f$ for different incident angles, comparing outputs from the designed structure (solid lines) with the ideal outputs assuming perfect transmission and perfect Strehl ratio (dashed lines).
Lens parameters: NA = 0.9, FOV = 60$^{\circ}$, $D_{\rm out}=50\lambda$, $D_{\rm in}=D_{\rm in}^{\rm opt} = 25\lambda$, $h=h_{\rm min}^{\rm Li\&Hsu}=5\lambda$.
}
\label{fig:Fig3}
\end{figure*}

During the optimizations, we update the permittivity profile with a macropixel size of $4\Delta x=\lambda/10$ to reduce the dimension of the design space and to keep the minimal feature size large while performing the simulations with a finer resolution of $\Delta x=\lambda/40$ for accuracy.
Since the desired response is symmetric, we impose $\varepsilon_{\rm r}({\bf r})$ to be mirror-symmetric with respect to the lens center $y=0$ and only parameterize $\varepsilon_{\rm r}$ of the macropixels in the left half of the metalens [red box in Fig.~\ref{fig:Fig3}(a)] as the optimization variables~\cite{2023_Li_ACSph}.

\section{Validation of theoretical bounds}\label{sec:bound_vs_design}

Before integrating theoretical bounds into the design process, it is prudent to first
verify whether the bounds we use are valid and if they are tight.
Such a validation is nontrivial and requires inverse design, since intuition-based designs are not likely to reach these bounds.
For this validation, we consider a relatively small system: metalenses with NA = 0.9, FOV = 60$^{\circ}$, and output diameter $D_{\rm out}=16\lambda$. 

In Ref.~\cite{li2023transmission}, we derived that the transmission efficiency averaged over all input channels within the FOV, $\langle T \rangle$, cannot exceed $N_{\rm eff}/N_{\rm in}$ where $N_{\rm in}$ is the number of input channels, and the inverse participation ratio $N_{\rm eff}=\left( \sum_i \sigma_i^2\right) ^2 /\sum_i \sigma_i^4$ quantifies the effective number of high-transmission channels using the singular values $\{\sigma_i\}$ of the transmission matrix.
After writing down the ideal transmission matrix of diffraction-limited wide-FOV metalenses in air (Supplementary Sec.~2), we plot $N_{\rm eff}/N_{\rm in}$ as the black solid lines in Fig.~\ref{fig:Fig2}.
The efficiency bound $N_{\rm eff}/N_{\rm in}$ depends on the diameter $D_{\rm in}$ of the input aperture and has a local maximum at $D_{\rm in}=D_{\rm in}^{\rm opt} \approx 8\lambda$.

To verify this dependence, we perform the topology optimization of Eq.~(\ref{eq:opt_problem}) for different $D_{\rm in}$,
with the refractive index of each macropixel bounded within $[n_L,n_H]=[1.0,2.0]$ 
and with the metalens thickness being $h=2\lambda$. 
The green circles in Fig.~\ref{fig:Fig2}(a) show the highest $\langle I_a \rangle_a=\langle {\rm SR}\cdot T \rangle$ among 100 optimizations with different initial guesses; they agree strikingly well with $N_{\rm eff}/N_{\rm in}$ and exhibit the predicted local maximum at $D_{\rm in}=D_{\rm in}^{\rm opt}$, indicating that this efficiency bound of Ref.~\cite{li2023transmission} is both valid and tight.

Note that $N_{\rm eff}/N_{\rm in}$ is the predicted maximal $\langle T \rangle$ with the Strehl ratio fixed at ${\rm SR}_a = 1$ for all incident angles, while the inverse design here allows ${\rm SR}_a$ to vary and optimizes the worst-case ${\rm SR}\cdot T$.
Therefore, the optimized $\langle {\rm SR}\cdot T \rangle$ should only be compared to $N_{\rm eff}/N_{\rm in}$ when the optimized Strehl ratio is high. 
For small values of $D_{\rm in}$ here, none of the optimizations reached a high Strehl ratio, presumably because incident light in the overly small aperture $D_{\rm in}$ cannot spread enough to cover the much larger $D_{\rm out}$ to yield the ideal output.

Next, we consider two bounds on the device thickness~\cite{li2022thickness,2023_Miller_Science}.
In Ref.~\cite{li2022thickness}, Li and Hsu used the lateral spreading $\Delta W$ of spatially localized inputs to quantify the degree of nonlocality, bounding the thickness through an empirical relation $h \ge h_{\rm min}^{\rm Li\&Hsu} = \Delta W$.
In Ref.~\cite{2023_Miller_Science}, Miller considered a transverse aperture that divides the system and used the number $C$ of channels that cross the transverse aperture to quantify the degree of nonlocality, bounding the thickness with a diffraction heuristics $h \ge h_{\rm min}^{\rm Miller} = (C\lambda)/\left[2(1-\cos{\theta_{\rm max})n_H} \right]$. 
Here, we take the ``maximal internal angle'' to be $\theta_{\rm max}=90^{\circ}$ since an inhomogeneous refractive index profile can scatter light to all possible angles; using a smaller $\theta_{\rm max}$ will increase $h^{\rm Miller}$.
With the lens parameters here ($D_{\rm out}=16\lambda$, NA = 0.9, and FOV = 60$^{\circ}$), the Li \& Hsu bound yields $h_{\rm min}^{\rm Li\&Hsu} = 1.7\lambda$, and the Miller bound yields $h_{\rm min}^{\rm Miller} = 4\lambda$. 
These bounds are shown as vertical dot-dashed lines in Fig.~\ref{fig:Fig2}(b).
Note that the value of $h_{\rm min}^{\rm Miller}$ also implicitly depends on how the channel number $C$ is counted, as detailed in Supplementary Sec.~4.

The green circles in Fig.~\ref{fig:Fig2}(b) show the highest optimized $\langle {\rm SR}\cdot T \rangle$ among 100 initial guesses with $D_{\rm in}=D_{\rm in}^{\rm opt} = 8\lambda$.
The optimized efficiency is low when the thickness is smaller than $h_{\rm min}^{\rm Li\&Hsu}$. 
When the thickness goes above $h_{\rm min}^{\rm Li\&Hsu}$, 
the optimized $\langle {\rm SR}\cdot T \rangle$ converges toward $N_{\rm eff}/N_{\rm in}$, validating both the efficiency bound and the thickness bound. 
In this example, $h_{\rm min}^{\rm Li\&Hsu}$ marks the transition between low-efficiency and high-efficiency designs, while $h_{\rm min}^{\rm Miller}$ coincides with the thickness at which the efficiency bound $N_{\rm eff}/N_{\rm in}$ is reached.

\section{High-NA high-efficiency metalens guided by theoretical bounds}\label{sec:metalens}

Having verified the theoretical bounds, we now integrate them into the design process.
We still consider NA = 0.9 and FOV = 60$^{\circ}$, but increase the system size to $D_{\rm out}=50\lambda$ since parameter scanning is no longer needed.
In the first step, we choose $D_{\rm in}=D_{\rm in}^{\rm opt} = 25\lambda$, which maximizes the efficiency bound $N_{\rm eff}/N_{\rm in}$. 
In the second step, we choose the necessary thickness.
For this system size, we find $h_{\rm min}^{\rm Li\&Hsu} = 5 \lambda$ and $h_{\rm min}^{\rm Miller} = 4.5 \lambda$ (Supplementary Sec.~4),
and we use the larger one $h = h_{\rm min}^{\rm Li\&Hsu} = 5 \lambda$ for the inverse design.
Finally, in the third step, we perform topology optimization.
Here, we additionally include a regularizer in the objective function of the optimization to promote a binary design with $\sqrt{\varepsilon_{\rm r}(\mathbf{r})}$ being either $n_L$ or $n_H$ (Supplementary Sec.~5).
We launch 100 optimizations with random initial guesses and take the best case.

Supplementary Video~1 shows the evolution of the metalens structure, its Strehl ratio and transmission efficiency as a function of the incident angle, and the intensity profiles at the focal plane as the optimization progresses.
Figure~\ref{fig:Fig3} shows the final configuration and performance.
The Strehl ratio and transmission efficiency both exhibit a flat distribution across the target FOV,
with $\langle T \rangle = 98\%$, $\langle {\rm SR}\rangle = 92\%$, and $\langle {\rm SR}\cdot T \rangle = 90\%$. 
A tight, near-ideal, focus with a full-width at half maximum (FWHM) of around $0.55\lambda$
is achieved for all incident angles within the FOV [Fig.~\ref{fig:Fig3}(c-d)].
Supplementary Video~2 plots the full intensity profile and field profile for all incident angles.

\begin{figure*}[t]
\centering
\includegraphics[width=1\textwidth]{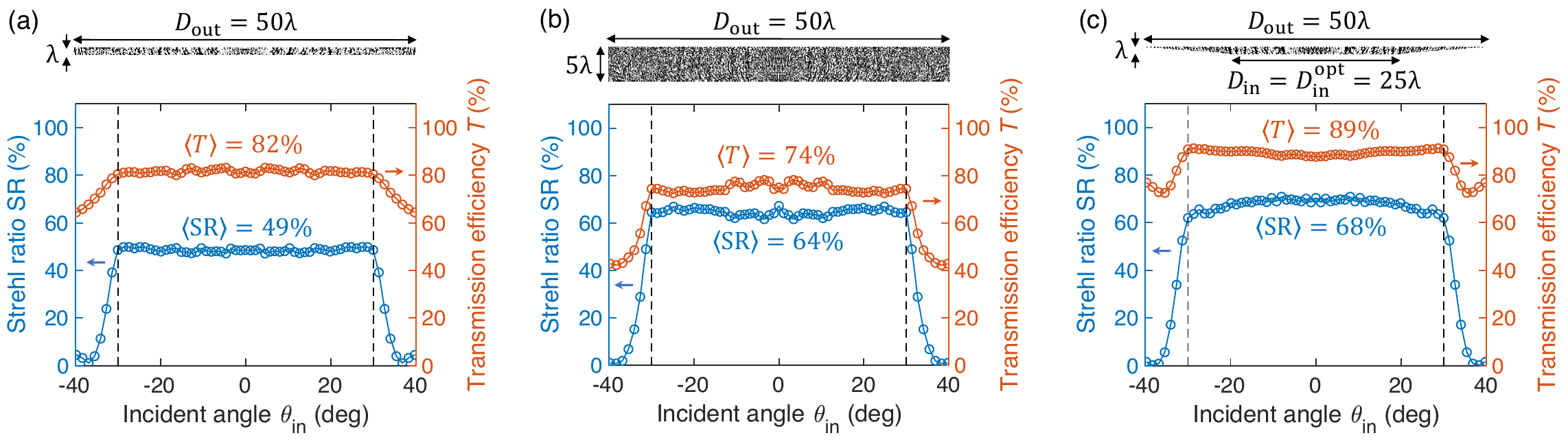}
\caption{Inverse-designed metalenses without guidance by theoretical bounds.
The lens parameters and optimization procedure are identical to those of Fig.~\ref{fig:Fig3} except that $D_{\rm in}=D_{\rm out}=50\lambda$ in (a-b) and the thickness is reduced to $h=\lambda$ in (a,c).
}
\label{fig:Fig4}
\end{figure*}

Integration of the theoretical bounds into the design process is crucial in enabling this optimal performance.
Previous inverse designs all adopted the common choice of $D_{\rm in}=D_{\rm out}$.
Figure~\ref{fig:Fig4}(a-b) shows the optimized results---following identical optimization steps---when we use $D_{\rm in}=D_{\rm out} = 50\lambda$ with $h=\lambda$ and $h=5\lambda$.
The optimized transmission efficiency $\langle T \rangle$ and Strehl ratio $\langle {\rm SR} \rangle$ are both greatly reduced.
The thickness choice is also critical.
Figure~\ref{fig:Fig4}(c) shows the optimized results when a typical thickness $h=\lambda$ is used together with $D_{\rm in}=D_{\rm in}^{\rm opt} = 25\lambda$, which is better than with $D_{\rm in}=D_{\rm out}$ but still fall substantially below the optimal performance of Fig.~\ref{fig:Fig3} where the thickness bound is adopted.

Given the non-convexity of the optimization problem, the optimized result depends on the initial guess, which is chosen randomly here.
Supplementary Fig.~S7 shows a histogram of the final $\langle {\rm SR}\cdot T \rangle$ among the 100 optimization runs for the four geometries in Figs.~\ref{fig:Fig3}--\ref{fig:Fig4}.

While the combination of Strehl ratio and transmission efficiency offers a comprehensive set of metrics, their measurement requires high-resolution large-area detectors and is not convenient. 
Therefore, a more commonly reported metric is the focusing efficiency, defined
as the ratio between the transmitted flux within three FWHM around the focal peak and the incident flux~\cite{banerji2019imaging,wang2022manipulation,sang2022toward,kim2020doublet,huang2021achromatic}.
We choose not to optimize such a focusing efficiency here because doing so may encourage the optimization to expand the FWHM and capture more light within the inflated 3$\times$FWHM, which increases the focusing efficiency under such definition but lowers the actual focusing quality.
To facilitate comparison with prior designs, we evaluate the focusing efficiency of the design in Fig.~\ref{fig:Fig3} (whose FWHM of $0.55\lambda$ is diffraction-limited); the angular dependence is shown in Supplementary Fig.~S6.
The focusing efficiency here averages to 88.5\% over the 60$^{\circ}$ FOV, which is higher even compared to the normal-incidence focusing efficiency of previous narrow-FOV metalenses (see Supplementary Tab.~S2).
We also evaluate the ideal focusing efficiency by taking the product of the transmission efficiency bound $N_{\rm eff}/N_{\rm in}$ and the focusing efficiency of an ideal focus with unity transmission; such an upper bound on the focusing efficiency is shown as solid lines in Fig.~\ref{fig:Fig1}(b) for $D_{\rm in}=D_{\rm out}$ and $D_{\rm in}=D_{\rm in}^{\rm opt}$.
The design in Fig.~\ref{fig:Fig3} saturates the maximized upper bound.

\section{Outlook}\label{sec:outlook}
This work shows that theoretical bounds can play an active role in photonic design, identifying optimal system parameters and maximizing the efficiency limit itself.
The maximizing-upper-limit strategy here can help a wide range of photonic devices reach their highest possible performance.

\section*{Acknowledgments}
Computing resources are provided by the Center for Advanced Research Computing at the University of Southern California.
\textbf{Funding}: This work was supported by the National Science Foundation CAREER award (ECCS-2146021) and the Sony Research Award Program.
\textbf{Disclosures}: The authors declare no conflicts of interest.
\textbf{Data Availability Statement}: All data needed to evaluate the conclusions in this study are presented in the paper and supporting information. The code is available on GitHub~\cite{inverse_design_code}.

\bibliography{apssamp}

\end{document}


\maketitle
\tableofcontents

\section{Angular spectrum propagation (ASP)}\label{sec:ASP}

We use angular spectrum propagation (ASP) to obtain the field profile after the metalens in the free space, which is an exact method equivalent to the Rayleigh-Sommerfeld diffraction integral.

For transverse magnetic (TM) waves at wavelength $\lambda$ in a 2D system, after obtaining the total field on the exit surface of the metalens $E_x(y,z=0)$ through full-wave simulations, we evaluate its Fourier components by
\begin{equation}
    E_x(k_y,z=0) = \frac{1}{\sqrt{2\pi}}\int_{-\infty}^{\infty}dy\ E_x(y,z=0)e^{-ik_yy},
    \label{eq:F_component}
\end{equation}
and then propagate it forward with
\begin{equation}
    E_x(k_y,z\ge 0) = E_x(k_y,z=0)e^{ik_zz}.
    \label{eq:propagate}
\end{equation}
Fourier transforming $E_x(k_y,z\ge 0)$ from the momentum basis to the spatial basis yields the field profile anywhere after the metalens
\begin{equation}
    E_x(y,z\ge 0) = \frac{1}{\sqrt{2\pi}}\int_{-\infty}^{\infty}dk_y\ E_x(k_y,z\ge 0)e^{ik_yy}.
    \label{eq:basis_convert}
\end{equation}

The continuous integration over the transverse momentum $k_y$ in Eq.~(\ref{eq:basis_convert}) can be approximated with a discrete summation.
A sampling with $2\pi/D_{\rm out}$ at the Nyquist rate should be sufficient since the field beyond the output aperture with diameter $D_{\rm out}$ is negligibly small.
However, this will introduce an artificial periodic boundary with periodicity $D_{\rm out}$ and produce periodic wrapping artifacts at the boundary.
Thus we use a finer grid of $k_y$ spaced by $2\pi/D_{\rm ASP}$ with $D_{\rm ASP}=2D_{\rm out}$.
Because we are not interested in the near-field region of the metasurface, we only consider the propagating Fourier components of $E_x(y,z\ge0)$ over $|k_y|<2\pi/\lambda$.

We perform full-wave simulations using MESTI~\cite{lin2022fast,MESTI} for the response of metalenses as $\mathbf{t}=\mathbf{CA}^{-1}\mathbf{B}$, where matrix $\mathbf{A}$ is the discretized Maxwell operator.
When matrices $\mathbf{B}$ and $\mathbf{C}$ contain plane waves at varied angles for source and projection profiles, the output is the angular transmission matrix $\mathbf{t}(k_y,k_y')$ detailed in Sec.~\ref{sec:ideal_TM}.
The total field $E_x(y,z=0)$ on the surface immediately after the metalens is evaluated as
\begin{equation}
    \frac{E_x^a(y,z=0)}{E_0} = \sum_bt_{ba}u_b(y),
    \label{eq:field_at_z_equal_0}
\end{equation}
where $t_{ba}$ is the transmission coefficient from the incident angle $\theta_{\rm in}^a$ to the outgoing angle $\theta_{\rm out}^b$; $u_b=\left(1/\sqrt{D_{\rm out}k_z^b}\right)e^{ik_y^by}$ is flux-orthogonal truncated plane waves with $\{ k_y^b \}=\left\{ b(2\pi/D_{\rm out})\text{ such that }b\in \mathbb{Z} \text{ and }|k_y^b|<2\pi/\lambda \right\}$ and $(k_y^b)^2+(k_z^b)^2=(2\pi/\lambda)^2$; constant $E_0$ represents the incident field. 
Integrating Eqs.~(\ref{eq:F_component})--(\ref{eq:field_at_z_equal_0}) into the output-projection matrix $\mathbf{C}$ directly yields the total field $E_x(y,z=f;\theta_{\rm in}^a)$ at the focal plane $z=f$. After extracting the field amplitude $E_x(y=f\tan{\theta_{\rm in}^b},z=f;\theta_{\rm in}^a)$ for all incident angles within the FOV, the diagonal elements $t_{aa}$ of the generalized transmission matrix are the focal spot amplitudes $E_x(y=f\tan{\theta_{\rm in}^a},z=f;\theta_{\rm in}^a)$.

\section{Transmission matrix of ideal wide-FOV metalens}\label{sec:ideal_TM}

\begin{figure}[t]
\centering
\includegraphics[width=0.6\textwidth]{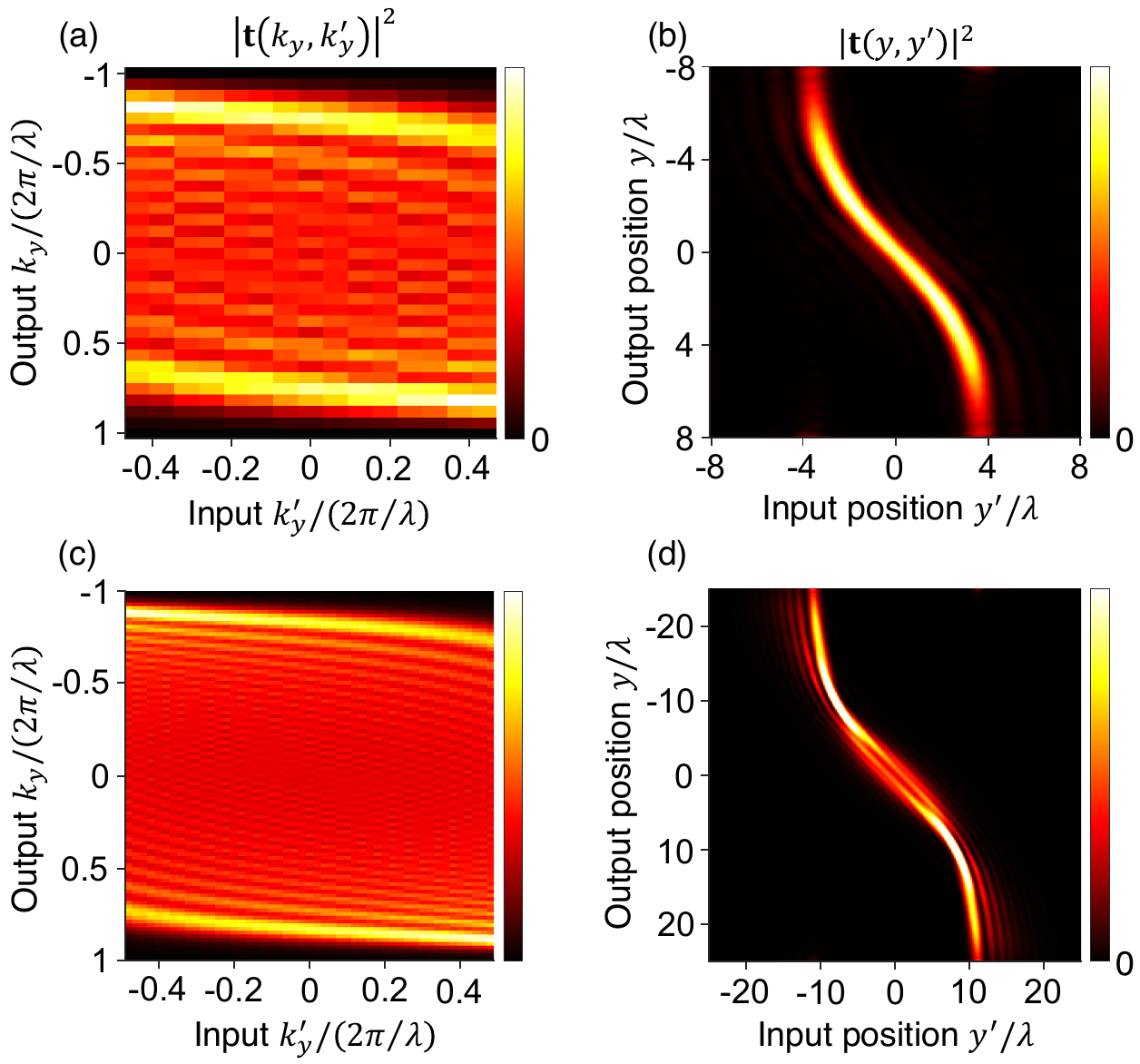}
\caption{Transmission matrices (a,c) in the angular basis $|\mathbf{t}(k_y,k_y')|^2$ and (b,d) in the spatial basis $|\mathbf{t}(y,y')|^2$. Lens parameters: NA = 0.9, FOV = 60$^{\circ}$, (a-b) output diameter $D_{\rm out}=16\lambda$, (c-d) output diameter $D_{\rm out}=50\lambda$.}
\label{fig:trans_matrix}
\end{figure}

To ideally focus a plane wave incident from angle $\theta_{\rm in}^a$ to point $(y=f\tan{\theta_{\rm in}^a},z=f)$ on the focal plane, the field on the exit surface of a metalens should be proportional to the conjugated field radiated from a point source at the focal spot, as
\begin{equation}
    E_x(y,z=0;\theta_{\rm in}^a)=\begin{cases}
 \frac{A(\theta_\mathrm{in}^a)e^{i\phi_\mathrm{out}^\mathrm{ideal}(y,\theta_\mathrm{in}^a)}}{[f^2+(y-f\tan{\theta_\mathrm{in}^a})^2]^{{1}/{4}}} & \text{for } |y|<\frac{D_\mathrm{out}}{2}, \\
    0 & \text{otherwise}.
    \end{cases}
    \label{eq:Ex_at_z_equal_0_ideal}
\end{equation}
where $A(\theta_{\rm in}^a)$ specifies the transmitted amplitude for incident angle $\theta_{\rm in}^a$ (in this work, $A(\theta_{\rm in}^a)$ is chosen for unitary transmission efficiencies across all incident angles within the FOV). 
The ideal phase distribution is
\begin{equation}
    \phi_{\rm out}^{\rm ideal}(y,\theta_{\rm in}^a) = \psi(\theta_{\rm in}^a)-\frac{2\pi}{\lambda}\sqrt{f^2+(y-f\tan{\theta_{\rm in}^a})^2},
    \label{eq:ideal_phase}
\end{equation}
where the angle-dependent global phase $\psi(\theta_{\rm in}^a)$ has no effect on the focusing quality; here $\psi(\theta_{\rm in}^a)=(2\pi/\lambda)\langle \sqrt{f^2+(y-f\tan{\theta_{\rm in}^a})^2}+y\sin{\theta_{\rm in}^a} \rangle_y$ is chosen to center phase profiles at different incident angles around the same $y$-averaged phase, and $\langle \cdots \rangle_y$ denotes averaging over $y$ within $|y|<D_{\rm out}/2$.

The angular transmission matrix $\mathbf{t}(k_y^b,k_y^a)=t_{ba}$ relating the incident wave from angle $\theta_{\rm in}^a$ to the outgoing wave $\theta_{\rm out}^b$ can be obtained from
\begin{equation}
    t_{ba} = \sqrt{\frac{k_z^b}{D_{\rm out}}}\int_{-\frac{D_{\rm out}}{2}}^{\frac{D_{\rm out}}{2}}E_x(y,z=0;\theta_{\rm in}^a)e^{-ik_y^by}dy \approx \Delta y\sqrt{\frac{k_z^b}{D_{\rm out}}}e^{-ik_y^b(-\frac{D_{\rm out}}{2}+\frac{\Delta y}{2})}\sum_{n=0}^{N-1}E_x(y_n,z=0;\theta_{\rm in}^a)e^{-i\frac{2\pi}{N}bn},
\end{equation}
where $y\in [-D_{\rm out}/2,D_{\rm out}/2]$ is discretized to $\{ y_n \equiv -D_{\rm out}/2+\Delta y/2+n\Delta y$ and $N\equiv D_{\rm out}/\Delta y \in \mathbb{Z}$.
Converting the transmission matrix from the momentum space to the spatial space yields
\begin{equation}
    \mathbf{t}(y,y') = 
    \begin{cases}
        \frac{1}{\sqrt{D_{\rm in}D_{\rm out}}}\sum_{b,a}\frac{k_z^a}{k_z^b}t_{ba}e^{-ik_y^by'} & \text{ when }|y|<D_{\rm out}/2 \text{ and } |y'|<D_{\rm in}/2 \\
        0 & \text{ otherwise}
    \end{cases}.
\end{equation}
Here, the input momentum satisfies $\{ k_y^a \}=\left\{ a(2\pi/D_{\rm in})\text{ such that }a\in \mathbb{Z} \text{ and }|k_y^a|<(2\pi/\lambda)\sin{(\rm FOV/2)} \right\}$ and $(k_y^a)^2+(k_z^a)^2=(2\pi/\lambda)^2$;
the output momentum satisfies $\{ k_y^b \}=\left\{ b(2\pi/D_{\rm out})\text{ such that }b\in \mathbb{Z} \text{ and }|k_y^b|<2\pi/\lambda \right\}$ and $(k_y^b)^2+(k_z^b)^2=(2\pi/\lambda)^2$;
a finer sampling of $\Delta y'=\frac{\lambda}{20\sin{(\rm FOV/2)}}$ and $\Delta y=\lambda/20$ is adopted for the input and output respectively such that transmission matrices of a small metalens have sufficient elements to ensure an accurate estimate of bounds.

The ideal transmission matrices in the angular and spatial bases adopted in the main text are shown in Fig.~\ref{fig:trans_matrix}. 

\section{Gradient obtained via adjoint method}\label{sec:gradient}

As introduced in the main text, the focusing quality is quantified by the ratio between the focal spot intensity of an actual design and that for an ideal metalens, as
\begin{equation}
    I_a = \frac{|E_x(y=f\tan{\theta_{\rm in}^a},z=f;\theta_{\rm in}^a)|^2}{\max_y|E_{x,{\rm ideal}}(y,z=f;\theta_{\rm in}^a)|^2} = \frac{|\mathbf{C}_a\mathbf{A}^{\rm -1}\mathbf{B}_a|^2}{\max_y|E_{x,{\rm ideal}}(y,z=f;\theta_{\rm in}^a)|^2}.
    \label{eq:I_a}
\end{equation}
The actual focal intensity can be mapped out with respect to the incident angle through the generalized transmission matrix $\mathbf{t}=\mathbf{CA}^{\rm -1}\mathbf{B}$, where the discretized Maxwell differential operator $\mathbf{A}$ is related to the relative permittivity profile $\varepsilon_{\rm r}(\mathbf{r})$ of the metalens; the input matrix $\mathbf{B}=[\mathbf{B}_1,\cdots,\mathbf{B}_M]$ contains the source profile that generates a truncated plane wave within the input aperture at the $a$-th incident angle $\theta_{\rm in}^a$; the output matrix $\mathbf{C}=[\mathbf{C}_1;\cdots;\mathbf{C}_M]$ propagates the output field of the metalens to the desired focal position on the focal plane $(y=f\tan{\theta_{\rm in}^a},z=f)$ by angular spectrum propagation (see Sec.~\ref{sec:ASP}).

Since the input and output matrices $\mathbf{B}$ and $\mathbf{C}$ are independent of $\varepsilon_{\rm r}({\mathbf{r}})$, using the identity $\partial \mathbf{A}^{-1}/\partial \varepsilon_{\rm r}(\mathbf{r}) = -\mathbf{A}^{-1}(\partial \mathbf{A}/\partial \varepsilon_{\rm r}(\mathbf{r}))\mathbf{A}^{-1}$,
we get
\begin{equation}
    \frac{dI_a}{d\varepsilon_{\rm r}(\mathbf{r})}=-\frac{2{\rm Re}\left[ (\mathbf{C}_a\mathbf{A}^{-1}\mathbf{B}_a)^*\mathbf{C}_a\mathbf{A}^{-1}\frac{\partial \mathbf{A}}{\partial \varepsilon_{\rm r}(\mathbf{r})}\mathbf{A}^{-1}\mathbf{B}_a \right]}{\max_y|E_{x,{\rm ideal}}(y,z=f;\theta_{\rm in}^a)|^2}
    \label{eq:dI_depsilon_1}
\end{equation}
with $^*$ representing the complex conjugation.
Under finite-difference discretization with grid size $\Delta x$,
the partial derivative of $\mathbf{A}$ with respect to a single permittivity component yields a matrix with a single non-zero entry, as
\begin{equation}
    \left[ \frac{\partial \mathbf{A}}{\partial \varepsilon_{\rm r}(\mathbf{r}_i)} \right]_{jk}=
    \begin{cases}
        -k_0^2\Delta x^2, & j=k=i \\
        0, & \text{otherwise} \\
    \end{cases},
    \label{eq:pA_pepsilon_1}
\end{equation}
where $k_0=2\pi/\lambda$ is the free-space wave vector. Thus
\begin{equation}
    \frac{\partial \mathbf{A}}{\partial \varepsilon_{\rm r}(\mathbf{r})}\mathbf{A}^{-1}\mathbf{B}_a = -k_0^2\Delta x^2 \cdot {\rm diag}(\mathbf{A}^{-1}\mathbf{B}_a),
    \label{eq:pA_pepsilon_2}
\end{equation}
and diag$(\mathbf{A}^{-1}\mathbf{B}_a)$ returns a square diagonal matrix with the elements of vector $\mathbf{A}^{-1}\mathbf{B}_a$ on the main diagonal.
Plugging Eq.~(\ref{eq:pA_pepsilon_2}) into Eq.~(\ref{eq:dI_depsilon_1}), the gradient of $I_a$ with respect to the permittivity profile can then be written as
\begin{equation}
    \frac{dI_a}{d\varepsilon_{\rm r}(\mathbf{r})}=\frac{2k_0^2\Delta x^2{\rm Re}\left[ (\mathbf{C}_a\mathbf{A}^{-1}\mathbf{B}_a)^*(\mathbf{A}^{-1}\mathbf{C}_a^{\rm T}\circ\mathbf{A}^{-1}\mathbf{B}_a) \right](\mathbf{r})}{\max_y|E_{x,{\rm ideal}}(y,z=f;\theta_{\rm in}^a)|^2},
    \label{eq:dI_depsilon_2}
\end{equation}
where $\circ$ denotes the element-wise product between two vectors. 

\section{Thickness bounds}\label{sec:bound}

\begin{figure}[t]
\centering
\includegraphics[width=0.55\textwidth]{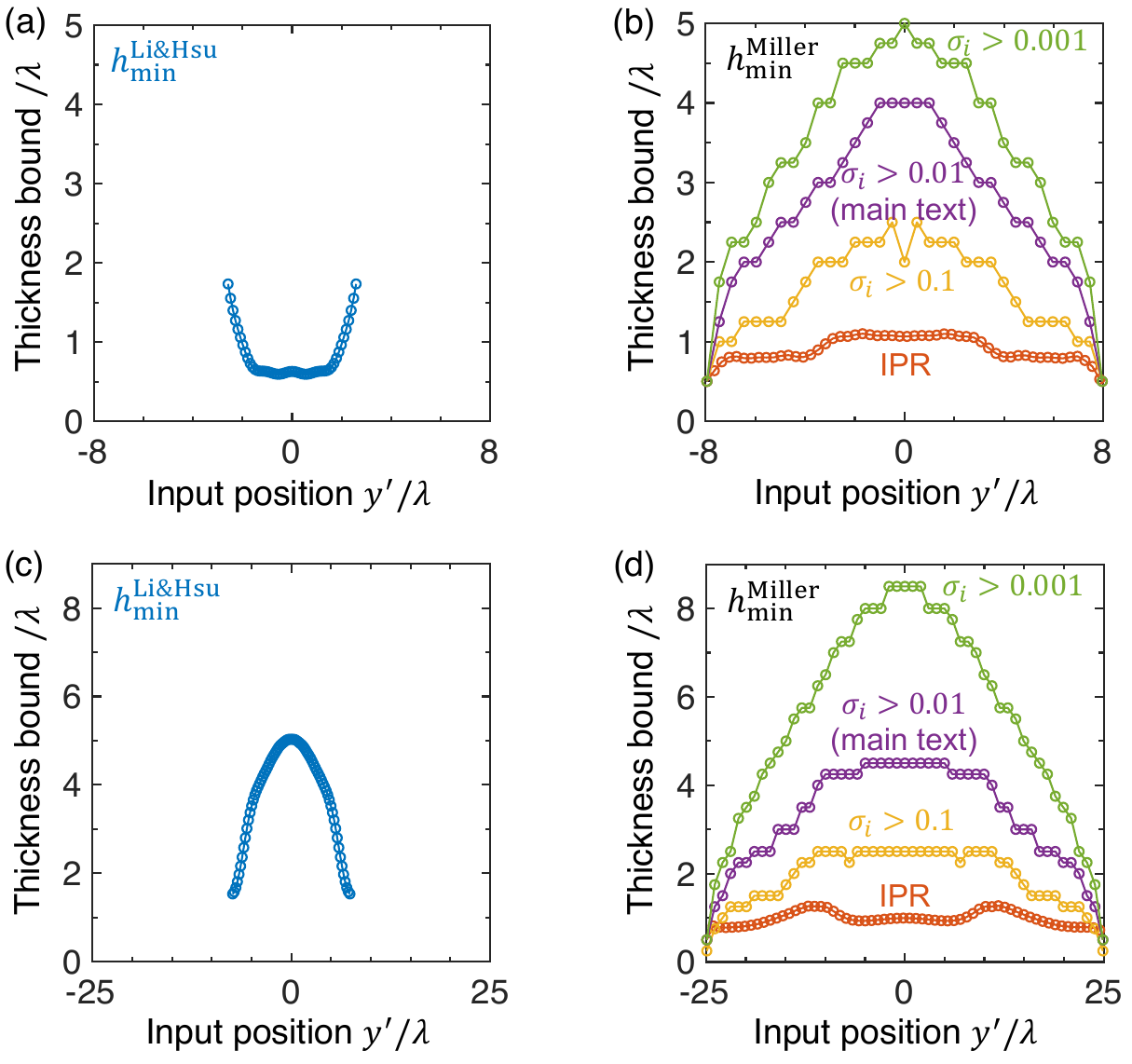}
\caption{Thickness bounds from Li \& Hsu~\cite{li2022thickness} and from Miller~\cite{2023_Miller_Science} applied to a wide-FOV lens.
Lens parameters: NA = 0.9, FOV = 60$^{\circ}$, (a,b) output diameter $D_{\rm out}=16\lambda$, (c,d) output diameter $D_{\rm out}=50\lambda$.
See text for the different definitions on the number of strongly coupled singular values used to establish $h_{\rm min}^{\rm Miller}$.
}
\label{fig:bound_vs_y}
\end{figure}

After writing down the ideal transmission matrices in Fig.~\ref{fig:trans_matrix}, we can obtain the minimal required thickness by evaluating the degree of nonlocality in the transmission matrix following Refs.~\cite{li2022thickness,2023_Miller_Science}.

Li and Hsu~\cite{li2022thickness} quantified the degree of nonlocality through the lateral spreading, defined as
\begin{equation}
    \Delta W(y') = W_{\rm out}(y') - W_{\rm in},
    \label{eq:lateral_spreading}
\end{equation}
where $W_{\rm out}(y')=[\int|\mathbf{t}(y,y')|^2dy]^2/\int|\mathbf{t}(y,y')|^4dy$ is the output beam width after passing through the structure with thickness $h$, quantified by the inverse participation ratio (IPR); $\mathbf{t}(y,y')$ is the transmission matrix in spatial basis shown in Fig.~\ref{fig:trans_matrix}, with a localized input at $y'$ on the input surface and a localized detection at $y$ on the output surface;
$W_{\rm in}=3\lambda/[4\sin{(\rm FOV/2)}]$ is the IPR width of the incident beam in the spatial transmission matrix, which consists of plane waves with momenta $|k_y'|<(2\pi/\lambda)\sin{(\rm FOV/2)}$.
An empirical relation established from full-wave simulations, 
\begin{equation}
    h \ge h_{\rm min}^{\rm Li\&Hsu} = \Delta W,
    \label{eq:Li_bound}
\end{equation}
links the lateral spreading to the minimal thickness.
Figure~\ref{fig:bound_vs_y}(a,c) shows the lateral spreading $\Delta W$ as a function of $y'$.
We confine our evaluation of $\Delta W(y')$ to the range of $y'$ where the output width $W_{\rm out}(y')$ of the spatial transmission matrix is clearly defined (see Fig.~\ref{fig:trans_matrix}). 
Taking the maximal lateral spreading in this range, we obtain $h_{\rm min}^{\rm Li\&Hsu}=1.7\lambda$ for an output diameter of $D_{\rm out}=16\lambda$ and $h_{\rm min}^{\rm Li\&Hsu}=5\lambda$ when $D_{\rm out}=50\lambda$ for NA = 0.9 and FOV = 60$^{\circ}$. 

Miller~\cite{2023_Miller_Science} quantified the degree of nonlocality through the total number of channels $C$ passing through a virtual transverse aperture at $y'$ that divides the structure into two sides: the left (L) side $y<y'$ and the right (R) side $y>y'$.
This transverse aperture at $y'$ divides the spatial transmission matrix into four blocks, and we consider the off-diagonal blocks
$\mathbf{t}_{\rm RL}$ and $\mathbf{t}_{\rm LR}$ that characterize light transmission from the left side of the input to the right side of the output, and from the right side of the input to the left side of the output.
The number $C$ of crossing channels is defined as $C=C_{\rm RL}+C_{\rm LR}$, where $C_{\rm RL}$ and $C_{\rm LR}$ are the number of strongly coupled singular values $\{\sigma_i\}$ of $\mathbf{t}_{\rm RL}$ and $\mathbf{t}_{\rm LR}$ respectively.
A minimal thickness is needed to support this number of crossing channels, quantified through a diffraction heuristics,
\begin{equation}
    h \ge h_{\rm min}^{\rm Miller} = \frac{C\lambda}{2(1-\cos{\theta_{\rm max}})n_H},
    \label{eq:Miller_bound}
\end{equation}
where $\theta_{\rm max}$ is the maximal internal angle and $n_H=2$ is the maximal refractive index.
Here, we adopt $\theta_{\rm max}=90^{\circ}$ since the inhomogeneous refractive index profile we use can scatter light to all possible angles; using a smaller $\theta_{\rm max}$ will increase $h^{\rm Miller}$.
Since the number of crossing channels depends on the location $y'$ of the transverse aperture, we scan $h^{\rm Miller}$ as a function of $y'$ and take its maximal value, as shown in Fig.~\ref{fig:bound_vs_y}(b,d). 
Note that the number of strongly coupled singular values depends on how they are counted.
Following Ref.~\cite{2023_Miller_Science}, we normalize the singular values $\{\sigma_i\}$ by the largest singular value at $y'=0$ and define $C_{\rm RL}$ and $C_{\rm LR}$ as the number of normalized singular values above a threshold.
Figure~\ref{fig:bound_vs_y}(b,d) shows the resulting $h_{\rm min}^{\rm Miller}$ for three relative thresholds: 0.001, 0.01, and 0.1.
In the main text, we use the version with relative threshold 0.01, which is the choice used in Ref.~\cite{2023_Miller_Science}.
Given the sensitive dependence on this threshold, here we also consider defining $C_{\rm RL}$ and $C_{\rm LR}$ by the IPR: $C_{\rm RL(LR)}=\left( \sum_i \sigma_i^2 \right)^2/\sum_i\sigma_i^4$, with the resulting $h_{\rm min}^{\rm Miller}$ shown in Fig.~\ref{fig:bound_vs_y}(b,d).
In the wide-FOV metalens example here, counting $C_{\rm RL}$ and $C_{\rm LR}$ with a relative threshold of 0.01 yields $h_{\rm min}^{\rm Miller}=4\lambda$ for $D_{\rm out}=16\lambda$ and $h_{\rm min}^{\rm Miller}=4.5\lambda$ for $D_{\rm out}=50\lambda$. 
Counting $C_{\rm RL}$ and $C_{\rm LR}$ with the IPR yields $h_{\rm min}^{\rm Miller}=1.1\lambda$ for $D_{\rm out}=16\lambda$ and $h_{\rm min}^{\rm Miller}=1.3\lambda$ for $D_{\rm out}=50\lambda$.

\section{Binarization}\label{sec:binary}

The gray-scale optimization problem described in Eq.~(2) and used in Fig.~2 of the main text allows the permittivity of each pixel to change continuously. 
For the optimizations for Fig.~3 and Fig.~4, we add a gradually turned-on regularizer to the objective function, which allows easier exploration of the design space at the beginning while enforcing binarization of permittivities $\varepsilon_{\rm r}(\mathbf{r})$ to either $n_L^2$ or $n_R^2$ as the optimization progresses.
To do so, we add a penalty term $P$ to the objective function, with~\cite{sell2019adjoint,phan2019high}
\begin{equation}
P = \frac{\|\rho(\mathbf{r})-0.5\|^2}{\|\rho_{\rm binary}(\mathbf{r})-0.5\|^2} \times
\begin{cases}
\left( \frac{m}{400} \right)^2, & m<400, \\
1, & m \ge 400,
\end{cases}
\label{eq:regularizer}
\end{equation}
where $m$ is the iteration number; $\rho(\mathbf{r})$ is a vector of normalized permittivity at all pixels of the design region with $\varepsilon_{\rm r}(\mathbf{r})=\rho(\mathbf{r})n_H^2+\left[ 1-\rho(\mathbf{r}) \right]n_L^2$; $\rho_{\rm binary}(\mathbf{r})\in \{0,1\}$ is a binarized permittivity profile used for normalization, and $\|\cdots\|$ denotes the vector norm.

\section{Evolution of the optimization}\label{sec:evol_opt}

\begin{figure}[t]
\centering
\includegraphics[width=0.8\textwidth]{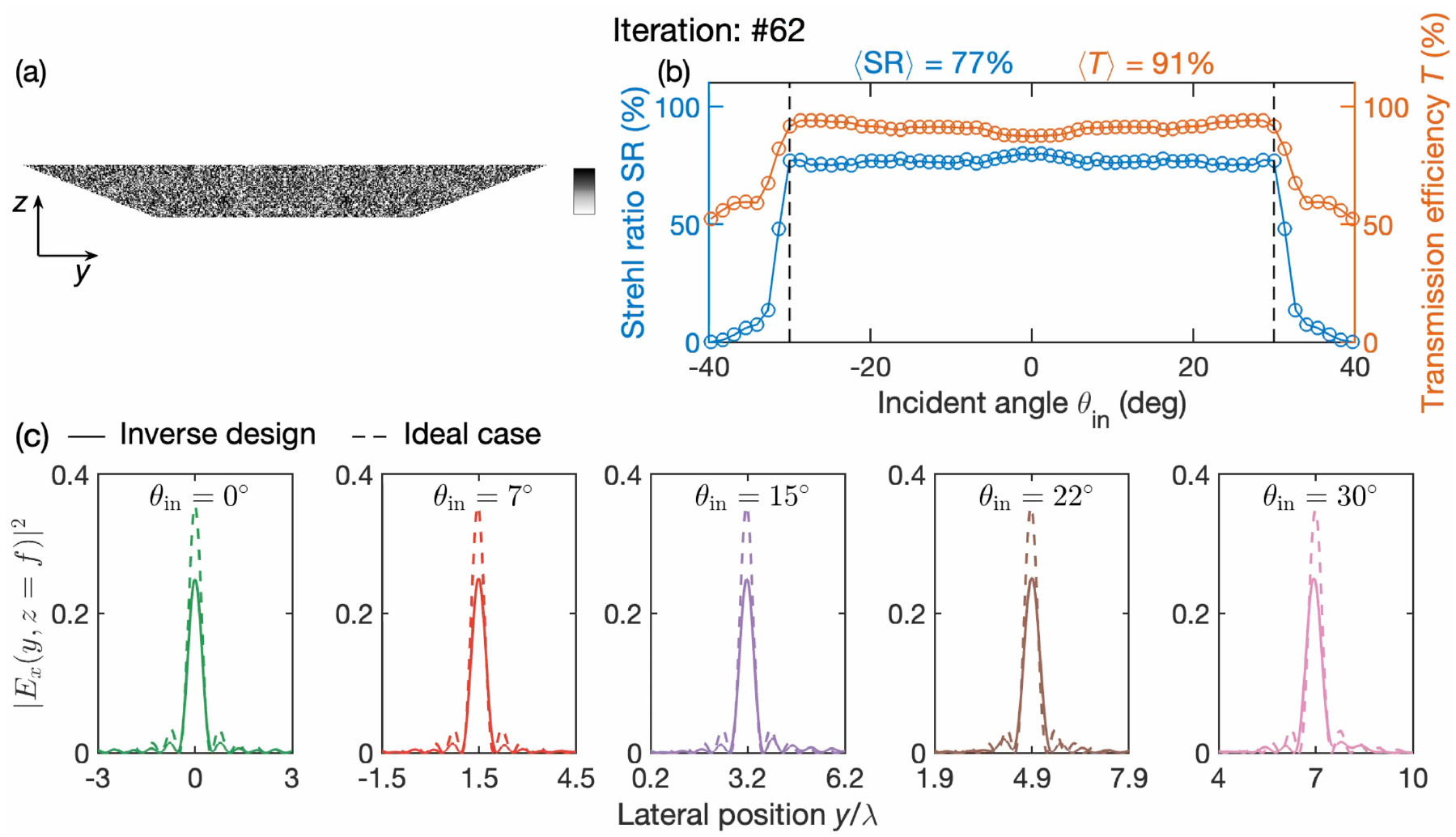}
\caption{One frame (at the 62-th iteration) of Supplementary Video~1, which shows how (a) the metalens structure, (b) its Strehl ratio and transmission efficiency, and (c) the focal intensity profile evolve as the optimization progresses.
Here, $\langle \cdots \rangle$ in (b) denotes averaging over incident angles within the FOV. Solid and dashed lines in (c) represent the intensity distribution of actual and ideal metalenses respectively.}
\label{fig:opt_animate}
\end{figure}

\begin{figure}[tb]
\centering
\includegraphics[width=0.4\textwidth]{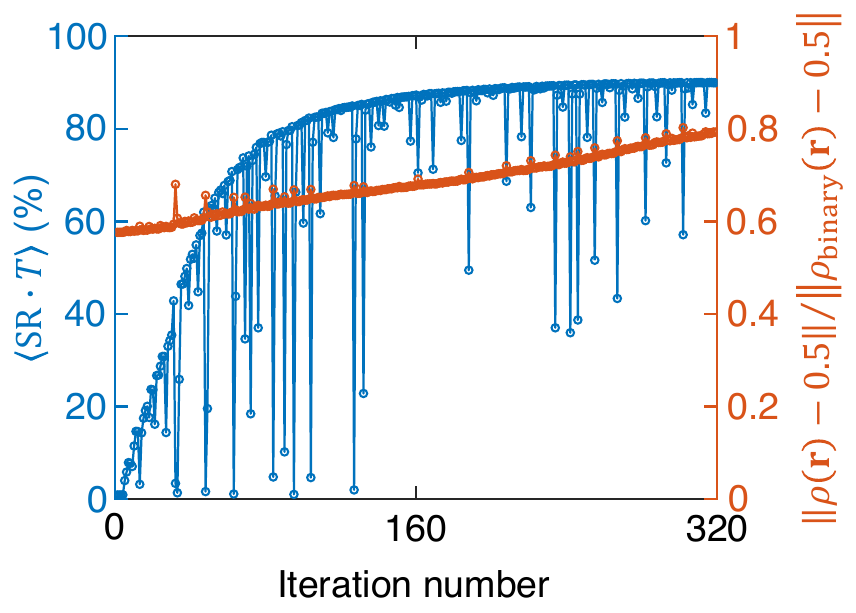}
\caption{Angle-averaged multiplication of the Strehl ratio and transmission efficiency over a 60$^{\circ}$ field of view and the binarization strength $\| \bm{\rho}-0.5 \|/\| \bm{\rho}_{\rm binary}-0.5 \|$ evolve as the optimization progresses.}
\label{fig:FoM_n_binary}
\end{figure}

For the best case shown in Fig.~3 of the main text ($D_{\rm out}=50\lambda$, $D_{\rm in}=25\lambda$, $h=5\lambda$),
Supplementary Video~1 shows how the metalens structure, its Strehl ratio, transmission efficiency, and intensity profile at the focal plane under different incident angles evolve as the optimization progresses.
Figure~\ref{fig:opt_animate} provides the animation caption and shows one frame of the animation.

Figure~\ref{fig:FoM_n_binary} plots the evolution of the angle-averaged $\langle {\rm SR}\cdot T \rangle$ and binarization strength $\| \rho(\mathbf{r})-0.5 \|/\| \rho_{\rm binary}(\mathbf{r})-0.5 \|$ as the optimization progresses. 

The optimization here took 13 hours and used 2.5 GiB of memory when running on a single core on an Intel Xeon 6130 node.

\section{Performance of the optimized metalens}\label{sec:FE}

\begin{figure}[t]
\centering
\includegraphics[width=0.42\textwidth]{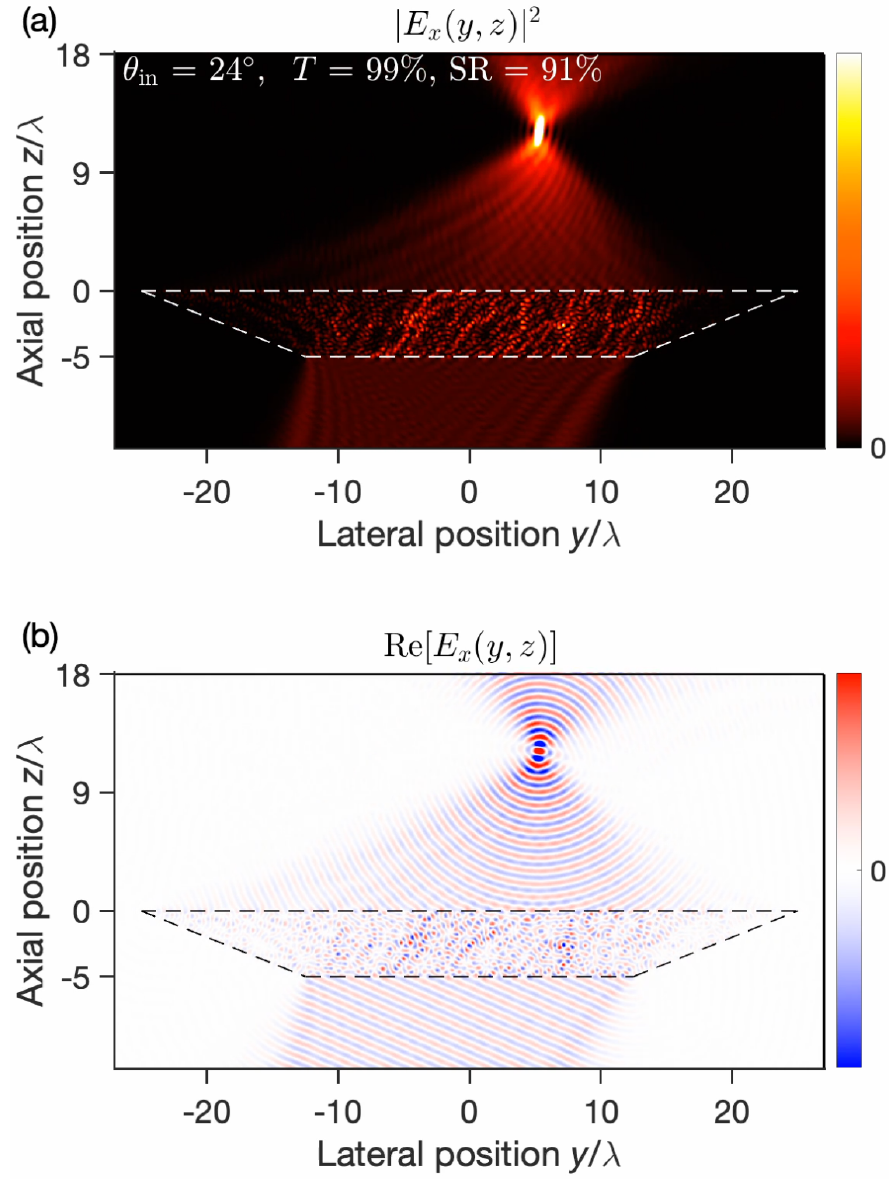}
\caption{One frame of Supplementary Video~2, which shows (a) $|E_x(y,z)|^2$ and (b) ${\rm Re}[E_x(y,z)]$ under plane-wave incidence from angle $\theta_{\rm in}=24^{\circ}$. The dashed lines mark the position of the metalens.}
\label{fig:focusing_animate}
\end{figure}

For the design shown in Fig.~3 of the main text ($D_{\rm out}=50\lambda$, $D_{\rm in}=25\lambda$, $h=5\lambda$),
Supplementary Video~2 shows $|E_x(y,z)|^2$ and ${\rm Re}[E_x(y,z)]$ across all angles within the FOV, with one frame of the animation provided in Fig.~\ref{fig:focusing_animate}.

\begin{figure}[h]
\centering
\includegraphics[width=0.3\textwidth]{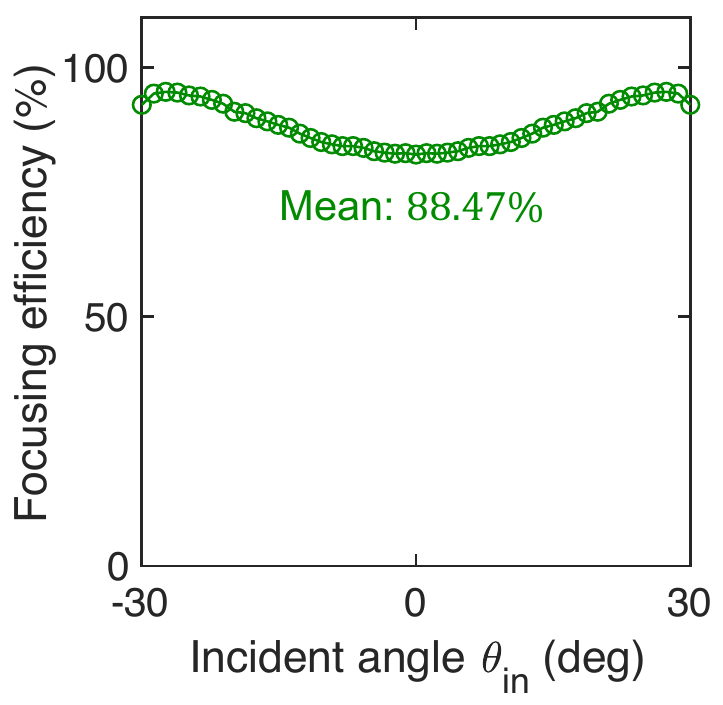}
\caption{The focusing efficiency of the optimal metalens design shown in Fig.~3 of the main text as a function of the incident angle $\theta_{\rm in}$, with an average of 88.47\%.}
\label{fig:FE_vs_angle}
\end{figure}

The focusing efficiency is defined as the ratio between the transmitted power within three full-widths at half-maximum (FWHM) around the focal peak and the incident power.
Figure~\ref{fig:FE_vs_angle} plots the incident-angle dependence of the focusing efficiency over the FOV, which averages to 88.47\%.

\section{Histogram of the optimized efficiency}\label{sec:histogram}

\begin{figure}[h]
\centering
\includegraphics[width=0.42\textwidth]{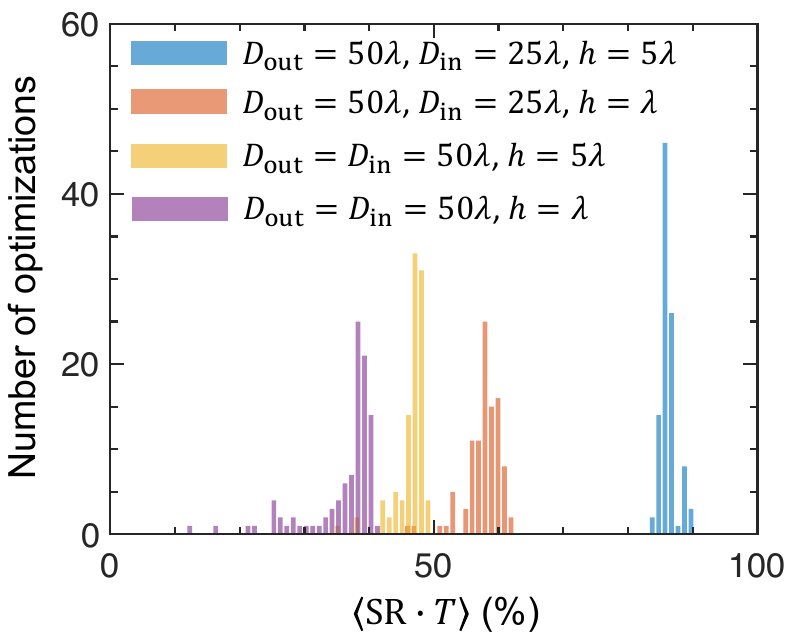}
\caption{Four groups of histograms for metalenses with different parameters, each showing the distribution of optimized $\langle {\rm SR}\cdot T \rangle$ across 100 optimizations from random initial guesses.}
\label{fig:opt_histogram}
\end{figure}

Figures~3--4 of the main text show the best results among optimizations with 100 random initial guesses, for different metalens parameters: output diameter $D_{\rm out}=50\lambda$, input diameter $D_{\rm in}=D_{\rm in}^{\rm opt}=25\lambda$, thickness $h=5\lambda$; $D_{\rm out}=50\lambda$, $D_{\rm in}=D_{\rm in}^{\rm opt}=25\lambda$, $h=\lambda$; $D_{\rm out}=D_{\rm in}=50\lambda$, $h=5\lambda$; $D_{\rm out}=D_{\rm in}=50\lambda$, $h=\lambda$.
Figure~\ref{fig:opt_histogram} plots the histogram of the final $\langle {\rm SR}\cdot T \rangle$ among the 100 instances for each set of system parameters.

\section{Literature survey of metalens designs}\label{sec:compare}

Table~\ref{tab:WFOV} lists the detailed information of wide-FOV metalens designs in the literature, which are summarized in Fig.~1(b) of the main text. 

Additionally, Table~\ref{tab:single_incidence} lists the information of narrow-FOV metalenses (designed for normal incidence only) with NA $\ge 0.9$.

\begin{table}[t]
\newcommand{\tabincell}[2]{\begin{tabular}{@{}#1@{}}#2\end{tabular}}
\centering
\setlength\extrarowheight{-3pt}
\caption{\bf Parameters and performance metrics of wide-field-of-view metalenses. ``-'' means not reported.}
\begin{tabular}{c|c|c|c|c|c|c|c|c}
\toprule
   \textbf{Exp./Sim.}$^a$ & \textbf{Wavelength} & \tabincell{c}{\textbf{Numerical} \\ \textbf{aperture}} & \textbf{FOV} & \textbf{Strehl ratio} & \textbf{FWHM} & \tabincell{c}{\textbf{Transmission} \\ \textbf{efficiency}} & \tabincell{c}{\textbf{Focusing} \\ \textbf{efficiency}} & \tabincell{c}{\textbf{Reference} \\ \textbf{(main text)}} \\
\midrule
    \multicolumn{9}{c}{\textbf{Monochromatic metalens}} \\
\midrule
    Exp. & 850 nm & 0.20 & 30$^{\circ}$ & - & - & - & Max: 20\% & [34] \\
\midrule
    Sim. & 940 nm & 0.20 & 180$^{\circ}$ & $\gtrsim0.8$ & - & - & 41\%--88\% & [35] \\
\midrule
    Exp. & 5.2 \textmu m & 0.24 & 180$^{\circ}$ & $\gtrsim0.8$ & - & - &  32\%--45\% & [35] \\
\midrule
    Sim. & 5 \textmu m & 0.24 & 180$^{\circ}$ & $\gtrsim0.8$ & - & - & 65\%--82\% & [39] \\
\midrule
    Sim. & 10.6 \textmu m & 0.24 & 140$^{\circ}$ &  $\gtrsim0.8$ & - & - & Avg: 50\% & [40] \\
\midrule
    Sim. & 532 nm & 0.25 & 170$^{\circ}$ & $\gtrsim0.8$ & - & - & 45\%--82\% & [36] \\
\midrule
    Exp. & 532 nm & 0.44 & 50$^{\circ}$ & $\gtrsim0.8$ & 1.1$\lambda$ & - & 30\%--50\% & [42] \\
\midrule
    Sim. & 632.8 nm & 0.45 & 50$^{\circ}$ & $\gtrsim0.5$ & 0.92$\lambda$ & - & Avg: 9\% & [45] \\
\midrule
    Exp. & 850 nm & 0.49 & 60$^{\circ}$ & $\gtrsim0.9$ & - & - & 45\%--70\% & [41] \\
\midrule
    Sim. & $\lambda$ & 0.90 & 60$^{\circ}$ & Avg: 0.92 & 0.55$\lambda$ & Avg: 98\% & Avg: 88\% & this work \\
\midrule
    \multicolumn{9}{c}{\textbf{Achromatic metalens}} \\
\midrule
    Sim. & 450--650 nm & 0.10 & 50$^{\circ}$ & $\gtrsim0.8$ & 5$\lambda$ & - & 38\%--62\% & [47] \\
\midrule
    Sim. & $0.8\lambda$--$\lambda$ & 0.12 & 16$^{\circ}$ & $\gtrsim0.8$ & - & - & Avg: 71\% & [48] \\
\midrule
    Sim. & 1--1.2 \textmu m & 0.18 & 180$^{\circ}$ & 0.64 & - & - & 10\%--80\% & [37] \\
\midrule
    Sim. & $0.8\lambda$--$\lambda$ & 0.24 & 60$^{\circ}$ & $\gtrsim0.8$ & - & - & Avg: 55\% & [48] \\
\midrule
    Sim. & 473/532 nm & 0.35 & 40$^{\circ}$ & - & 1.5$\lambda$ & - & 7\%--12\% & [43] \\
\midrule
    Sim. & 445/532/660 nm & 0.33/0.38/0.47 & 60$^{\circ}$ & - & - & - & 7\%--53\% & [44] \\
\midrule
    Sim. & 470--650 nm & 0.61 & 60$^{\circ}$ & - & 0.9$\lambda$ & - & 15\%--70\% & [46] \\
\midrule
    Sim. & $0.8\lambda$--$\lambda$ & 0.70 & 80$^{\circ}$ & $\gtrsim0.8$ & - & - & Avg: 25\% & [48] \\
\midrule
    Exp. & 33--37 GHz & 0.81 & 90$^{\circ}$ & - & 0.55$\lambda$ & - & 36\%--47\% & [38] \\
\bottomrule
\end{tabular}
  \label{tab:WFOV}
\end{table}

\begin{table}[t]
\newcommand{\tabincell}[2]{\begin{tabular}{@{}#1@{}}#2\end{tabular}}
\centering
\setlength\extrarowheight{-3pt}
\caption{\bf Parameters and performance metrics of narrow-FOV metalenses designed for normal incidence with NA $\ge 0.9$. ``-'' means not reported.}
\begin{threeparttable}
\begin{tabular}{c|c|c|c|c|c|c|c}
\toprule
   \textbf{Exp./Sim.} & \textbf{Wavelength} & \tabincell{c}{\textbf{Numerical} \\ \textbf{aperture}} & \textbf{Strehl ratio} & \textbf{FWHM} & \tabincell{c}{\textbf{Transmission} \\ \textbf{efficiency}} & \tabincell{c}{\textbf{Focusing} \\ \textbf{efficiency}} & \tabincell{c}{\textbf{Reference} \\ \textbf{(main text)}} \\
\midrule
    \multicolumn{8}{c}{\textbf{Monochromatic metalens}} \\
\midrule
    Exp. & 532 nm & 0.90 & 0.9 & 0.57$\lambda$ & - & 40\% & [19] \\
\midrule
    Sim. & 640 nm & 0.90 & 0.8 & - & - & 78\% & [30] \\
\midrule
    Sim. & 633 nm & 0.90 & - & - & 80\% & 40\% & [25] \\
\midrule
    Sim. & 500 nm & 0.90 & - & - & - & 56\%$\ ^a$ & [28] \\
\midrule
    Sim. & 580 nm & 0.94 & - & - & - & 79\% & [29] \\
\midrule
    Sim. & 633 nm & 0.95 & - & \tabincell{c}{0.76$\lambda$ \\ 0.44$\lambda$} & 76\% & 30\% & [25] \\
\midrule
    Sim. & 118.8 \textmu m & 0.95 & - & \tabincell{c}{0.78$\lambda$ \\ 0.42$\lambda$} & - & 66\% & [27] \\
\midrule
    Sim. & 1550 nm & 0.97 & - & 0.51$\lambda$ & 75\% & 55\% & [18] \\
\midrule
    Sim. & 1550 nm & 0.97 & - & 0.55$\lambda$ & - & 87\% & [31] \\
\midrule
    Sim. & 532 nm & 0.98 & - & 0.52$\lambda$ & - & 71\% & [20] \\
\midrule
    Sim. & 715 nm & 0.99 & - & 0.54$\lambda$ & - & 37\% & [21] \\
\midrule
    Sim. & 633 nm & 0.99 & - & - & 60\% & 14\% & [25] \\
\midrule
    \multicolumn{8}{c}{\textbf{Achromatic metalens}} \\
\midrule
    Sim. & 450--750 nm & 0.90 & - & - & - & Avg: 34\% & [22] \\
\midrule
    Sim. & 450--700 nm & 0.90 & - & 0.86$\lambda$--0.89$\lambda$ & - & Avg: 41\% & [23] \\
\midrule
    Sim. & 7.5--10 GHz & 0.93 & - & 0.54$\lambda$ & - & Max: 48\% & [24] \\
\midrule
    Sim. & 500--650 nm & 0.94 & - & - & - & 43\%--75\% & [29] \\
\midrule
    Sim. & 1014--1114 nm & 0.97 & - & 0.69$\lambda$--0.76$\lambda$ & - & Avg: 44\% & [26] \\
\midrule
    Sim. & 450--700 nm & 0.99 & - & 0.86$\lambda$--0.89$\lambda$ & - & Avg: 27\% & [23] \\
\bottomrule
\end{tabular}
\begin{tablenotes}
    \footnotesize
        \item[a] Here ${\rm SR}\cdot T=56\%$, which is comparable to the focusing efficiency.
\end{tablenotes}
\end{threeparttable}
  \label{tab:single_incidence}
\end{table}


\bibliography{sample}
